\documentclass[noshowpacs,amsmath,
twocolumn,
superscriptaddress,
8pt,
aps,prb]{revtex4-2}
\pdfoutput=1
\usepackage{setspace}
\usepackage[left]{lineno}
\usepackage{amsmath}
\usepackage{multirow}
\usepackage{graphicx}
\usepackage{upgreek}

\renewcommand{\figurename}{\textbf{Fig.}}
\usepackage{verbatim}
\usepackage{amsfonts}
\usepackage{amssymb}
\usepackage{epstopdf}
\usepackage{dcolumn}
\usepackage{xcolor}
\usepackage{verbatim}
\usepackage{siunitx}
\setcitestyle{super}
\DeclareGraphicsExtensions{.pdf,.eps,.png,.jpg,.mps}
\begin{document}
\title{Soliton microcombs in X-cut LiNbO$_3$ microresonators}

\author{Binbin Nie$^{1,\dagger}$, Xiaomin Lv$^{1,2,\dagger}$, Chen Yang$^{1,\dagger}$, Rui Ma$^{3}$, Kaixuan Zhu$^{1}$, Ze Wang$^{1}$, Yanwu Liu$^{1}$, Zhenyu Xie$^{1}$, Xing Jin$^{1}$, Guanyu Zhang$^{1}$, Du Qian$^{1}$, Zhenyu Chen$^{3}$, Qiang Luo$^{1,3}$, Shuting Kang$^{3}$, Guowei Lv$^{1,2,4,5}$, Qihuang Gong$^{1,24,5}$, Fang Bo$^{3,*}$, and Qi-Fan Yang$^{1,2,4,5,*}$\\
$^1$State Key Laboratory for Artificial Microstructure and Mesoscopic Physics and Frontiers Science Center for Nano-optoelectronics, School of Physics, Peking University, Beijing, 100871, China\\
$^2$Hefei National Laboratory, Hefei, 230088, China\\
$^3$Nankai University, Tianjin, 300071, China\\
$^4$Collaborative Innovation Center of Extreme Optics, Shanxi University, Taiyuan, 030006,  China\\
$^5$Peking University Yangtze Delta Institute of Optoelectronics, Nantong, 226010, China\\
$^{\dagger}$These authors contributed equally to this work.\\
$^{*}$Corresponding author: bofang@nankai.edu.cn; leonardoyoung@pku.edu.cn}

\begin{abstract}
Chip-scale integration of optical frequency combs, particularly soliton microcombs, enables miniaturized instrumentation for timekeeping, ranging, and spectroscopy. Although soliton microcombs have been demonstrated on various material platforms, realizing complete comb functionality on photonic chips requires the co-integration of high-speed modulators and efficient frequency doublers, features that are available in a monolithic form on X-cut thin-film lithium niobate (TFLN). However, the pronounced Raman nonlinearity associated with extraordinary light in this platform has so far precluded soliton microcomb generation. Here, we report the generation of transverse-electric-polarized soliton microcombs with a 25 GHz repetition rate in high-Q microresonators on X-cut TFLN chips. By precisely orienting the racetrack microresonator relative to the optical axis, we mitigate Raman nonlinearity and enable soliton formation under continuous-wave laser pumping. Moreover, the soliton microcomb spectra are extended to 350 nm with pulsed laser pumping. This work expands the capabilities of TFLN photonics and paves the way for the monolithic integration of fast-tunable, self-referenced microcombs.

\vspace{6pt}
\end{abstract}

\maketitle
\begin{figure*}[!t]
\centering
    \includegraphics[scale = 1.0]{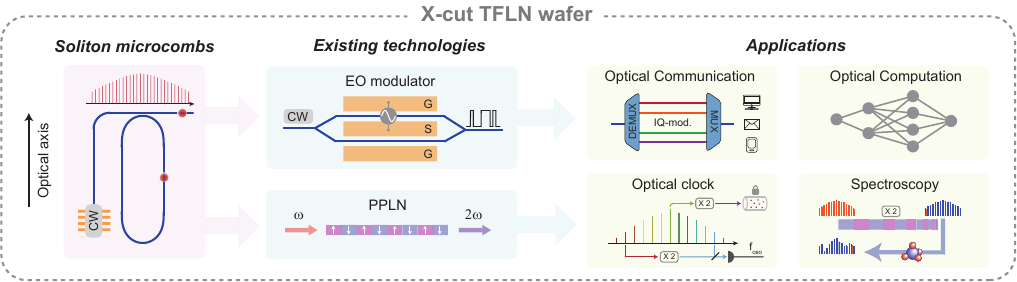}
    \caption{{ {\bf Technologies and applications based-on X-cut TFLN.}}
    EO: electro-optic; PPLN: periodically-poled lithium niobate; DEMUX: de-multiplexer; MUX: multiplexer.}
    \label{figure1}
\end{figure*}

\begin{figure*}[!ht]
\centering
\includegraphics[scale = 1.0]{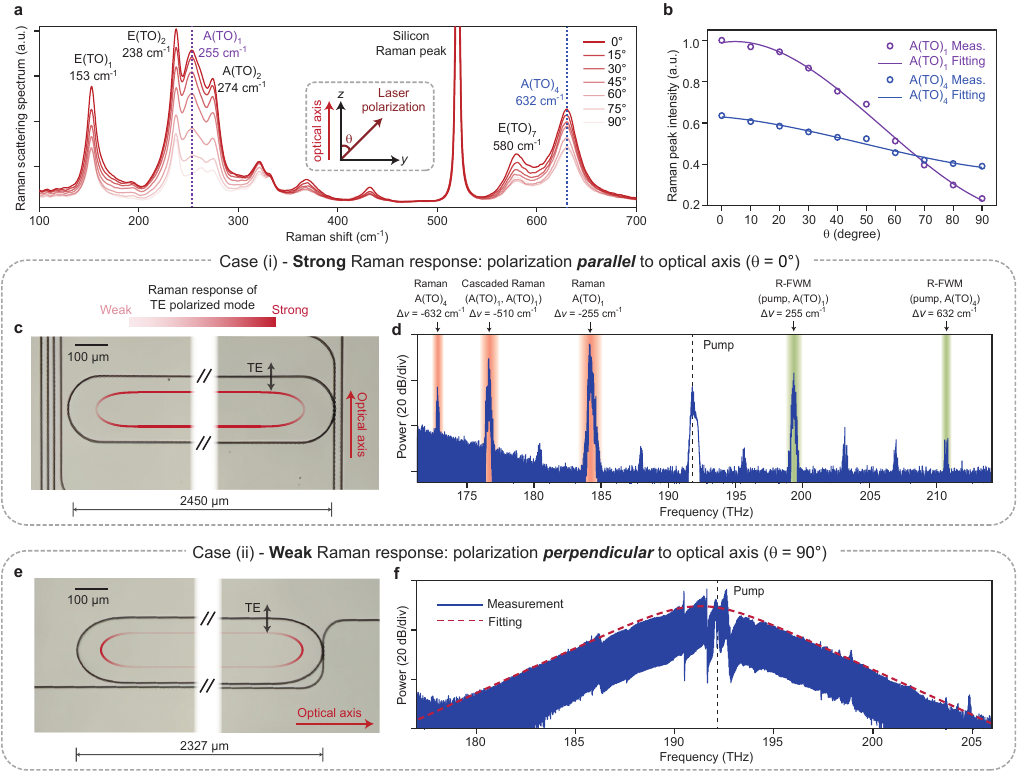}
\caption{{\bf Raman response of X-cut TFLN.}
\textbf{a,} Polarization-dependent Raman spectra of X-cut TFLN, with the angle between the pump laser polarization and the optical axis varying from 0$^{\circ}$ to 90$^{\circ}$. The vibrational modes are indicated.  
\textbf{b,} Raman peak intensity of vibrational modes A(TO)$_1$ and A(TO)$_4$ as a function of the angle between the pump laser polarization and the LN optical axis.  
\textbf{c,} Microscope image of LiNbO$_3$ racetrack microresonators, with the straight waveguide section oriented perpendicular to the optical axis. Color gradients indicate the local Raman response.  
\textbf{d,} Optical spectrum of Raman lasing and four-wave mixing generated in the microresonator shown in (c). The physical origin of each spectral component is labeled.
\textbf{e,} Microscope image of LiNbO$_3$ racetrack microresonators, with the straight waveguide section oriented parallel to the optical axis. Color gradients indicate the local Raman response.  
\textbf{f,} Optical spectrum of soliton microcomb generated in the microresonator shown in (e).
}  
\label{figure2}
\end{figure*}
\begin{figure}[!ht]
\centering
   \includegraphics[scale = 1.0]{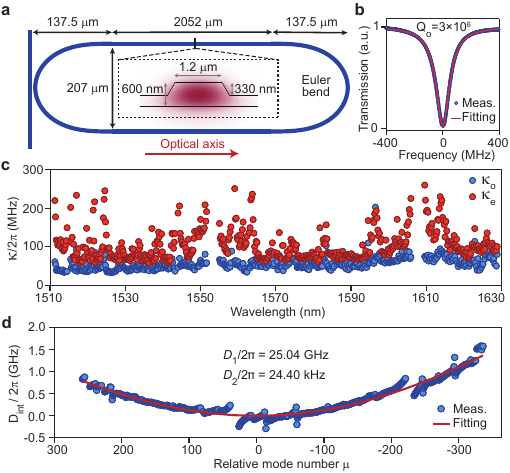}
   \caption{
   {\bf Design and properties of the microresonator.}
   \textbf{a,} Design of the microresonator (not to scale). The cross-sectional geometry with the profile of the TE fundamental mode is provided in the inset.
   \textbf{b,} Typical transmission spectrum of a fundamental TE mode. The intrinsic quality factor is derived from Lorentzian fitting as 3 million.
   \textbf{c,} The intrinsic ($\kappa_o$) and external ($\kappa_e$) coupling losses of the fundamental TE mode as a function of wavelength.
   \textbf{d,} Mode family dispersion of the fundamental TE mode. The integrated dispersion is defined as $D_\mathrm{int}(\mu)=\omega_\mu-\mu D_1=D_2\mu^2/2+\mathcal{O}(\mu^3)$ as a function of mode index $\mu$ relative to the pump. $\omega_\mu$ is the resonant frequency of the $\mu_\mathrm{th}$ mode and $D_1/2\pi$ is the FSR. The red line denotes parabolic fitting with $D_2/2\pi$=24.4 kHz.
   }
   \label{figure3}
\end{figure}
\begin{figure*}[!ht]
\centering
   \includegraphics[scale = 1.0]{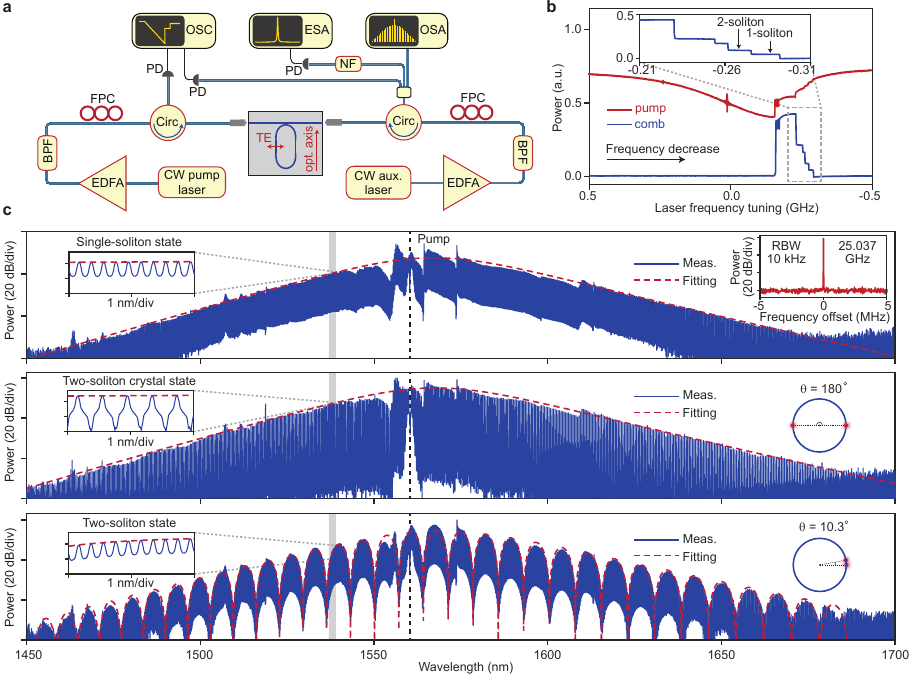}
   \caption{
   {\bf  Generation of soliton microcombs in X-cut LiNbO$_3$ microresonator.}  
   \textbf{a,} Experimental set-up. EDFA: erbium-doped fiber amplifier, BPF: band-pass filter, FPC: fiber polarization controller, Circ: circulator, PD: photodetector, NF: notch filter, OSC: oscilloscope, ESA: electrical spectrum analyzer, OSA: optical spectrum analyzer.
   \textbf{b,} Normalized transmitted pump power (red) and comb power (blue) as a function of pump laser frequency tuning. Inset: zoom-in view of the soliton steps.
   \textbf{c,} Optical spectra of single-soliton state, two-soliton crystal state, and two-soliton state. The dashed red lines represent fitted soliton spectral envelopes. The left insets show the zoom-in views of soliton comb lines. The electrical beatnote of the single soliton state is presented in the right upper inset. RBW: resolution bandwidth. For multi-soliton states, the relative angular positions of solitons are depicted in the corresponding right insets.
   }
   \label{figure4}
\end{figure*}
\begin{figure*}[!ht]
    \includegraphics[scale = 1.0]{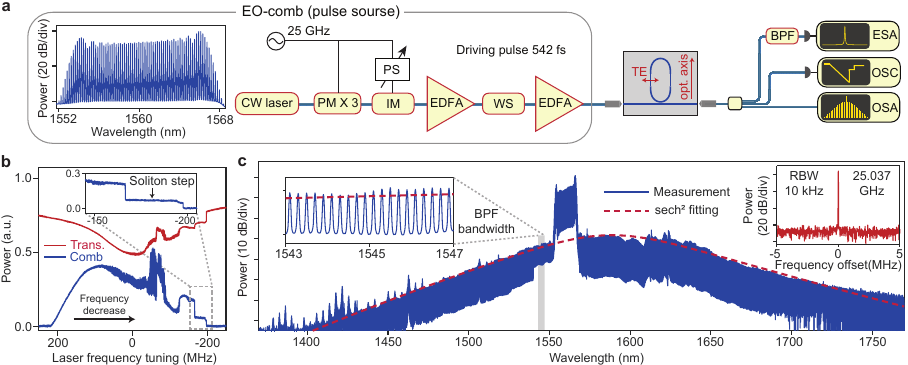}
    \caption{
    {\bf Pulse-pumped soliton microcombs.} 
    \textbf{a,} Experimental set-up. Left inset: optical spectrum of EO comb. PM: phase modulator; IM: intensity modulator; PS: phase shifter; WS: waveshaper.
    \textbf{b,} Normalized transmitted power (red) and comb power selected by the BPF (blue) as a function of pump laser frequency tuning. Inset: zoom-in view of the soliton step. 
    \textbf{c,} Optical spectrum of the soliton microcomb. The spectral envelope is fitted using sech$^2$ function (red dashed line). Left inset: zoom-in view of soliton comb lines within the BPF bandwidth. Right inset: electrical beanote of the comb lines selected by the BPF. RBW: resolution bandwidth.
    }
    \label{figure5}
\end{figure*}

{\bf\noindent Introduction}

The rapid evolution of integrated photonics is driving the demand for multifunctional material platforms that support diverse on-chip optical operations. Among these platforms, thin-film lithium niobate (TFLN) has emerged as a prime candidate owing to its ultralow optical losses, robust second-order nonlinear response ($d_{33} \approx -27\,\mathrm{pm/V}$), and high electro-optic efficiency ($\gamma_{33} \approx 30\,\mathrm{pm/V}$)~\cite{boes2023lithium,zhu2021integrated}. Traditionally employed for frequency doubling and electro-optic modulation, the performance of LiNbO$_3$ has been remarkedly enhanced by TFLN technology through improved confinement of both optical and electrical fields\cite{zhu2021integrated}. These advances have enabled high-speed modulators~\cite{wang2018integrated,xu2020high} and highly efficient frequency doublers~\cite{wang2018ultrahigh,lu2019periodically,chen2024adapted} that surpass previous capabilities (Fig.~\ref{figure1}). Notably, fully exploiting LiNbO$_3$’s maximum nonlinear coefficients and electro-optic properties requires that both the optical and electric fields be aligned parallel to the crystal’s optical axis -- a condition fulfilled by the transverse-electric (TE) modes in X-cut TFLN.

Chip-based optical frequency combs, or microcombs, have concurrently transformed the architectures of integrated photonics\cite{kippenberg2011microresonator,chang2022integrated}. These microcombs are indispensable for the co-integration of microwave and atomic systems in applications such as optical frequency synthesis\cite{spencer2018optical}, timekeeping\cite{newman2019architecture}, and advanced computational tasks\cite{feldmann2021parallel}. In TFLN, microcombs are generated via resonantly enhanced nonlinear optical processes -- most notably Kerr effects~\cite{he2019self,gong2020near,gong2019soliton,lv2024broadband} and electro-optic interactions~\cite{zhang2019broadband,rueda2019resonant,hu2022high} -- yielding comb lines described by $f_n = n f_r + f_o$, where $f_r$ is the repetition rate and $f_o$ is the carrier-envelope-offset frequency. Broadband microcombs spanning more than an octave have been demonstrated in Z-cut LiNbO$_3$ microresonators by balancing Kerr nonlinearity with anomalous dispersion, a requisite for soliton mode-locking \cite{wang2024octave,song2024octave}. However, the monolithic integration of soliton microcombs with efficient frequency doublers and high-speed electro-optic modulators is most straightforward on X-cut TFLN. In this configuration, the strong Raman response associated with extraordinary-polarized light has traditionally precluded soliton formation, instead favoring Raman lasing~\cite{schaufele1966raman,okawachi2017competition,lv2024broadband,wang2019monolithic,yu2020raman,gong2020photonic}.

In this work, we tailor the portion between ordinary and extraordinary light in LiNbO$_3$ microresonators to suppress stimulated Raman scattering on X-cut TFLN. Microwave-rate ($f_r\approx25$ GHz) soliton microcombs with 200 nm spectral span are generated in TE modes with a continuous-wave pump laser, whose spectra are further extended to over 350 nm using a pulsed pump laser. Our approach allows for seamless integration with existing technologies on X-cut TFLN, promising applications in optical communication\cite{marin2017microresonator}, computation\cite{feldmann2021parallel}, timing\cite{newman2019architecture}, and spectroscopy\cite{suh2016microresonator,bao2021architecture} (Fig.~\ref{figure1}).

\vspace{6pt}

{\bf\noindent Results}

{\bf\noindent Raman response of X-cut TFLN}

\noindent Firstly, we characterize the polarization-dependence of the Raman response of an X-cut TFLN chip using Raman spectroscopy\cite{schaufele1966raman,repelin1999raman} (Fig. \ref{figure2}a {  and Methods}). The Raman spectra contain multiple peaks corresponding to different vibrational modes, with intensities decreasing as the pump polarization transitions from parallel (extraordinary light) to perpendicular (ordinary light) to the optical axis. For instance, in the measurement, the peak intensities of vibrational modes A(TO)$_1$ and A(TO)$_4$ are reduced to 40\% and 20\%, respectively (Fig. \ref{figure2}b). Thus, similar to the case of LiTaO$_3$ \cite{wang2024lithium}, incorporating more ordinary light in the modes of LiNbO$_3$ microresonators can significantly mitigate the Raman effect, which is the key to our design.

To control the portion of ordinary light, we use racetrack microresonators instead of ring microresonators. The majority of the modes are confined in the straight waveguides, whose orientation determines the dominance of ordinary or extraordinary light. We test two racetrack microresonators with different orientations on X-cut TFLN-on-insulator chips. In device (i), the straight waveguides are perpendicular to the optical axis, such that the TE mode is polarized along the optical axis (Fig. \ref{figure2}c). Therefore, the Raman response of the TE mode is strong in the straight section but weak in the arc section. Measurements show that this device exhibits a loaded Q factor of approximately 1 million and anomalous group velocity dispersion, theoretically permitting soliton microcomb generation. However, the resulting optical spectrum shows Raman lasing due to interactions with vibrational modes A(TO)$_1$ and A(TO)$_4$ (Fig. \ref{figure2}d).

In contrast, in device (ii), the straight waveguides are parallel to the optical axis, and the TE mode is polarized perpendicular to the optical axis (Fig. \ref{figure2}e). As a result, the Raman response of the TE mode is weak in the straight section but strong in the arc section. With significantly reduced Raman nonlinearity, soliton microcombs are generated in this device, with the characteristic sech$^2$-shaped optical spectrum shown in Fig. \ref{figure2}f. Detailed characterization of the device and the soliton microcombs are presented in the following sections.

\vspace{6pt}
{\bf\noindent Device design and characterization}

\noindent The soliton-generating microresonators are fabricated from a 600-nm-thick LiNbO$_3$ layer atop a 4.7-µm-thick silica layer, following established procedures \cite{lv2024broadband}. The etching depth is 330 nm with an etching angle of 62 degrees (Fig. \ref{figure3}a). The waveguide's top width is 1.2 µm, supporting two TE modes in the straight sections and one TE mode in the arc sections. The 2052-µm-long straight waveguides are connected by Euler bends to suppress spatial mode interactions and reduce losses from the straight-to-bend transitions\cite{ji2022compact}. The free-spectral-range (FSR) of the microresonator is approximately 25 GHz. A bus waveguide with a top width of 1 µm is coupled to the microresonator at the Euler bend. The transmitted spectrum shows an intrinsic Q factor up to 3 million within the telecommunication C-band (Fig. \ref{figure3}b).

The wavelength-dependent leakage of the optical mode in the arc sections indicates larger coupling losses at longer wavelengths, a feature leveraged to suppress Raman lasing in LiNbO$_3$ microresonators \cite{he2019self,gong2020photonic,he2023high,lv2024broadband,song2024octave}. This is confirmed by the measured coupling and intrinsic losses of the fundamental TE mode shown in Fig. \ref{figure3}c. The Q factor exhibits abnormal degradation at wavelengths near 1560 nm and 1610 nm, attributed to coupling with the transverse-magnetic mode. This modal coupling is evident in the mode family dispersion (Fig. \ref{figure3}d), where notable deviations from parabolic fitting are observed at these wavelengths. Nevertheless, the TE mode predominantly follows anomalous group velocity, with a second-order dispersion ($D_2/2\pi$) of 24.4 kHz.

\vspace{6pt}
{\bf\noindent Soliton microcombs driven by continuous-wave lasers}

\noindent The experimental setup is illustrated in Fig. \ref{figure4}a, where a continuous-wave laser is amplified to pump the microresonator. The laser is tuned from the blue-detuned side to the red-detuned side of the mode to access the soliton state \cite{herr2014temporal}. This process is complicated by the thermo-optic effect, which red-shifts the mode with increasing intracavity power\cite{carmon2004dynamical}. This behavior contrasts with Z-cut TFLN microresonators, where photorefraction \cite{he2019self} and pyroelectricity \cite{zhang2025fundamental} counterbalance the thermo-optic effect. Consequently, direct tuning into the soliton state is infeasible due to the abrupt decrease in intracavity power \cite{carmon2004dynamical,yi2015soliton,brasch2016photonic}. To mitigate thermal nonlinearity, an auxiliary laser is tuned into a counter-propagating mode of the microresonator \cite{zhou2019soliton,zhang2019sub,lu2019deterministic}. The powers of the pump and auxiliary lasers on the bus waveguide are 192.5 mW and 148.9 mW, respectively. The existence range of the soliton in terms of the pump laser detuning is significantly increased with optimal auxiliary laser frequency (Fig. \ref{figure4}b). The discrete steps in the comb power represent comb states with distinct soliton numbers.

Figure \ref{figure4}c presents optical spectra of different soliton states, including single-soliton, two-soliton crystal, and two-soliton states. All spectra span more than 200 nm, with deviations from the spectral envelopes caused by mode-crossings. The electrical beat note of the single-soliton state, detected with a fast photodiode, exhibits a narrow-linewidth peak at 25.037 GHz, certifying the coherent nature of the soliton microcomb. For multi-soliton states, the envelope of the generated optical spectra shows pronounced modulation due to interference between individual solitons. The relative angular positions of solitons can be reconstructed by Fourier transforming the optical spectra, as depicted in the corresponding right insets.

\vspace{6pt}
{\bf\noindent Soliton microcombs driven by optical pulses}

\noindent Soliton microcombs can also be generated using synchronized pulsed lasers \cite{obrzud2017temporal}, which offer higher optical-to-optical conversion efficiencies \cite{li2022efficiency} and the potential to access a broader spectral range \cite{obrzud2019microphotonic,anderson2022zero,xiao2023near}. As illustrated in Fig. \ref{figure5}a, we synthesize an electro-optic (EO) comb from a continuous-wave laser using cascaded phase and intensity modulators \cite{murata2000optical}. Proper dispersion compensation with a programmable waveshaper generates transform-limited sinc-shaped pulses with a full width at half maximum (FWHM) of 542 fs. After amplification, an average power of 36.24 mW is launched into the bus waveguide. A bandpass filter is inserted at the output port to isolate the generated comb lines from the EO comb.

The repetition frequency of the EO comb is finely tuned around the microresonator's FSR of 25.037 GHz until the characteristic soliton step is observed during frequency scanning (Fig. \ref{figure5}b). Up to two solitons are supported under the pulsed laser pump, and the improved conversion efficiency allows for adiabatic access to the soliton state \cite{obrzud2017temporal}. The optical spectrum of the single soliton state spans from 1400 nm to 1750 nm, following a sech\(^2\)-shaped spectral envelope (Fig. \ref{figure5}c). Notably, a set of spurs deviating from the sech\(^2\) spectral envelope appears at wavelengths below 1470 nm. This deviation is attributed to birefringence-induced interaction between TE and TM modes, which occurs at shorter wavelengths \cite{wang2024lithium,zhang2025ultrabroadband} {  (Extended Data Fig. \ref{fig:ExtFig1})}. We use a bandpass filter centered at 1545 nm to select a portion of the generated comb lines for photodetection. The coherence of the generated comb lines is verified by their narrow-linewidth monotone electrical beat note.

\vspace{6pt}
\noindent\textbf{Discussion}

\noindent The demonstration of soliton microcomb on X-cut LiNbO$_3$ microresonators provides a clear route toward fully integrated on-chip comb functionality. Unlike Si$_3$N$_4$ microcombs, the X-cut LiNbO$_3$ platform enables monolithic integration with electrodes for high-speed modulation, thereby affording an additional degree of freedom for feedback control of both $f_r$ and $f_o$\cite{wang2018integrated,he2023high}. Furthermore, integration with PPLN waveguides \cite{wang2018ultrahigh, chen2024adapted} facilitates on-chip $f_o$ detection, contingent upon achieving an octave-spanning comb spectrum \cite{wang2024octave,song2024octave} -- a capability that necessitates both a higher power budget and meticulous dispersion engineering. These challenges can be overcome by employing coupled microresonators, which enhance comb generation efficiency while reducing power requirements \cite{xue2019super,hu2022high,helgason2023surpassing,yang2024efficient}.

The potential applications of microcombs on this platform are broad and compelling. For example, a monolithically integrated coherent photonic transceiver may employ the microcomb as a multi-wavelength light source in conjunction with arrays of high-speed modulators. To be compatible with current wavelength-division multiplexing systems, it is advantageous to set $f_r$ on the order of hundreds of gigahertz---a target more readily achieved with Kerr microcombs than with EO combs driven by microwave sources\cite{zhang2019broadband,hu2022high,yu2022integrated,zhang2025ultrabroadband}. Notably, our approach is also applicable to smaller racetrack microresonators with larger free-spectral ranges, wherein bending losses are substantially minimized by established deep-etching techniques \cite{gao2023compact} (Extended Data Fig. \ref{fig:ExtFig2}). Furthermore, high-power combs with flattened spectral envelopes can be realized in both dark-pulse\cite{xue2015mode,lv2024broadband} and dissipation-engineered configurations\cite{xue2023dispersion}. Finally, by incorporating chirped PPLN waveguides\cite{wu2024visible}, broadband comb spectra can be efficiently converted to the visible and mid-infrared regions for atomic and molecular spectroscopy \cite{suh2016microresonator,bao2021architecture}. Collectively, these advances lay the groundwork for the realization of chip-based optical clocks \cite{newman2019architecture}, building on recent progress in visible laser technologies\cite{lu2024emerging} and photonic-integrated atomic systems\cite{isichenko2023photonic,Hummon:18,stern2013nanoscale}.


\bigskip

\noindent\textbf{Methods}

\begin{footnotesize}

\noindent\textbf{Raman spectroscopy.} 
The Raman spectra presented in Fig. \ref{figure2}a are measured using Raman microscopes (LabRAM HR Evolution, Horiba). The polarization of the excitation laser is varied by rotating the half-wave plate positioned behind it, and the scattered Raman light is detected without polarization selection. Both the excitation laser and the detected Raman light propagate along the crystalline X axis, i.e. perpendicular to the chip. 
The intense peak at 520 cm$^{-1}$ is attributed to the Raman response of the thick silicon substrate of TFLN chip. In the notations used to label vibrational modes, TO refers to transverse optical phonons, while A and E denote non-degenerate and doubly degenerate vibrational modes, respectively.

\vspace{6pt}

\noindent\textbf{Device fabrication.} 
Devices were fabricated on a commercial X‐cut LiNbO\(_3\)-on-insulator wafer (NANOLN), which comprises a 600‐nm‐thick LiNbO\(_3\) layer atop a 4.7-\textmu m‐thick silica buffer layer. The patterns for the bus waveguides and microrings were defined via electron‐beam lithography on a 750‐nm‐thick layer of ma–N 2405 resist. These patterns were subsequently transferred onto the LiNbO\(_3\) film using Ar\(^+\) plasma etching, yielding an etch depth of 330 nm and sidewall angles of approximately 62\(^{\circ}\). Following the etching process, the photoresist was removed using organic solvents and oxygen plasma cleaning, and any residual redeposition was eliminated with an ammonia solution to ensure a pristine sample surface.
\vspace{6pt}

\end{footnotesize}
\medskip

\bibliography{ref}

\begin{thebibliography}{10}
\expandafter\ifx\csname url\endcsname\relax
  \def\url#1{\texttt{#1}}\fi
\expandafter\ifx\csname urlprefix\endcsname\relax\def\urlprefix{URL }\fi
\providecommand{\bibinfo}[2]{#2}
\providecommand{\eprint}[2][]{\url{#2}}

\bibitem{boes2023lithium}
\bibinfo{author}{Boes, A.} \emph{et~al.}
\newblock \bibinfo{title}{Lithium niobate photonics: Unlocking the electromagnetic spectrum}.
\newblock \emph{\bibinfo{journal}{Science}} \textbf{\bibinfo{volume}{379}}, \bibinfo{pages}{eabj4396} (\bibinfo{year}{2023}).

\bibitem{zhu2021integrated}
\bibinfo{author}{Zhu, D.} \emph{et~al.}
\newblock \bibinfo{title}{Integrated photonics on thin-film lithium niobate}.
\newblock \emph{\bibinfo{journal}{Adv. Opt. Photonics}} \textbf{\bibinfo{volume}{13}}, \bibinfo{pages}{242--352} (\bibinfo{year}{2021}).

\bibitem{wang2018integrated}
\bibinfo{author}{Wang, C.} \emph{et~al.}
\newblock \bibinfo{title}{Integrated lithium niobate electro-optic modulators operating at {CMOS}-compatible voltages}.
\newblock \emph{\bibinfo{journal}{Nature}} \textbf{\bibinfo{volume}{562}}, \bibinfo{pages}{101--104} (\bibinfo{year}{2018}).

\bibitem{xu2020high}
\bibinfo{author}{Xu, M.} \emph{et~al.}
\newblock \bibinfo{title}{High-performance coherent optical modulators based on thin-film lithium niobate platform}.
\newblock \emph{\bibinfo{journal}{Nat. Commun.}} \textbf{\bibinfo{volume}{11}}, \bibinfo{pages}{3911} (\bibinfo{year}{2020}).

\bibitem{wang2018ultrahigh}
\bibinfo{author}{Wang, C.} \emph{et~al.}
\newblock \bibinfo{title}{Ultrahigh-efficiency wavelength conversion in nanophotonic periodically poled lithium niobate waveguides}.
\newblock \emph{\bibinfo{journal}{Optica}} \textbf{\bibinfo{volume}{5}}, \bibinfo{pages}{1438--1441} (\bibinfo{year}{2018}).

\bibitem{lu2019periodically}
\bibinfo{author}{Lu, J.} \emph{et~al.}
\newblock \bibinfo{title}{Periodically poled thin-film lithium niobate microring resonators with a second-harmonic generation efficiency of 250,000\%/{W}}.
\newblock \emph{\bibinfo{journal}{Optica}} \textbf{\bibinfo{volume}{6}}, \bibinfo{pages}{1455--1460} (\bibinfo{year}{2019}).

\bibitem{chen2024adapted}
\bibinfo{author}{Chen, P.-K.} \emph{et~al.}
\newblock \bibinfo{title}{Adapted poling to break the nonlinear efficiency limit in nanophotonic lithium niobate waveguides}.
\newblock \emph{\bibinfo{journal}{Nat. Nanotechnol.}} \textbf{\bibinfo{volume}{19}}, \bibinfo{pages}{44--50} (\bibinfo{year}{2024}).

\bibitem{kippenberg2011microresonator}
\bibinfo{author}{Kippenberg, T.~J.}, \bibinfo{author}{Holzwarth, R.} \& \bibinfo{author}{Diddams, S.}
\newblock \bibinfo{title}{Microresonator-based optical frequency combs}.
\newblock \emph{\bibinfo{journal}{Science}} \textbf{\bibinfo{volume}{332}}, \bibinfo{pages}{555--559} (\bibinfo{year}{2011}).

\bibitem{chang2022integrated}
\bibinfo{author}{Chang, L.}, \bibinfo{author}{Liu, S.} \& \bibinfo{author}{Bowers, J.~E.}
\newblock \bibinfo{title}{Integrated optical frequency comb technologies}.
\newblock \emph{\bibinfo{journal}{Nat. Photon.}} \textbf{\bibinfo{volume}{16}}, \bibinfo{pages}{95--108} (\bibinfo{year}{2022}).

\bibitem{spencer2018optical}
\bibinfo{author}{Spencer, D.~T.} \emph{et~al.}
\newblock \bibinfo{title}{An optical-frequency synthesizer using integrated photonics}.
\newblock \emph{\bibinfo{journal}{Nature}} \textbf{\bibinfo{volume}{557}}, \bibinfo{pages}{81--85} (\bibinfo{year}{2018}).

\bibitem{newman2019architecture}
\bibinfo{author}{Newman, Z.~L.} \emph{et~al.}
\newblock \bibinfo{title}{Architecture for the photonic integration of an optical atomic clock}.
\newblock \emph{\bibinfo{journal}{Optica}} \textbf{\bibinfo{volume}{6}}, \bibinfo{pages}{680--685} (\bibinfo{year}{2019}).

\bibitem{feldmann2021parallel}
\bibinfo{author}{Feldmann, J.} \emph{et~al.}
\newblock \bibinfo{title}{Parallel convolutional processing using an integrated photonic tensor core}.
\newblock \emph{\bibinfo{journal}{Nature}} \textbf{\bibinfo{volume}{589}}, \bibinfo{pages}{52--58} (\bibinfo{year}{2021}).

\bibitem{he2019self}
\bibinfo{author}{He, Y.} \emph{et~al.}
\newblock \bibinfo{title}{Self-starting bi-chromatic {LiNbO$_3$} soliton microcomb}.
\newblock \emph{\bibinfo{journal}{Optica}} \textbf{\bibinfo{volume}{6}}, \bibinfo{pages}{1138--1144} (\bibinfo{year}{2019}).

\bibitem{gong2020near}
\bibinfo{author}{Gong, Z.}, \bibinfo{author}{Liu, X.}, \bibinfo{author}{Xu, Y.} \& \bibinfo{author}{Tang, H.~X.}
\newblock \bibinfo{title}{Near-octave lithium niobate soliton microcomb}.
\newblock \emph{\bibinfo{journal}{Optica}} \textbf{\bibinfo{volume}{7}}, \bibinfo{pages}{1275--1278} (\bibinfo{year}{2020}).

\bibitem{gong2019soliton}
\bibinfo{author}{Gong, Z.} \emph{et~al.}
\newblock \bibinfo{title}{Soliton microcomb generation at 2 \textmu m in z-cut lithium niobate microring resonators}.
\newblock \emph{\bibinfo{journal}{Opt. Lett.}} \textbf{\bibinfo{volume}{44}}, \bibinfo{pages}{3182--3185} (\bibinfo{year}{2019}).

\bibitem{lv2024broadband}
\bibinfo{author}{Lv, X.} \emph{et~al.}
\newblock \bibinfo{title}{Broadband microwave-rate dark pulse microcombs in dissipation-engineered {L}i{N}b{O}$_3$ microresonators}.
\newblock \emph{\bibinfo{journal}{arXiv preprint arXiv:2404.19584}}  (\bibinfo{year}{2024}).

\bibitem{zhang2019broadband}
\bibinfo{author}{Zhang, M.} \emph{et~al.}
\newblock \bibinfo{title}{Broadband electro-optic frequency comb generation in a lithium niobate microring resonator}.
\newblock \emph{\bibinfo{journal}{Nature}} \textbf{\bibinfo{volume}{568}}, \bibinfo{pages}{373--377} (\bibinfo{year}{2019}).

\bibitem{rueda2019resonant}
\bibinfo{author}{Rueda, A.}, \bibinfo{author}{Sedlmeir, F.}, \bibinfo{author}{Kumari, M.}, \bibinfo{author}{Leuchs, G.} \& \bibinfo{author}{Schwefel, H.~G.}
\newblock \bibinfo{title}{Resonant electro-optic frequency comb}.
\newblock \emph{\bibinfo{journal}{Nature}} \textbf{\bibinfo{volume}{568}}, \bibinfo{pages}{378--381} (\bibinfo{year}{2019}).

\bibitem{hu2022high}
\bibinfo{author}{Hu, Y.} \emph{et~al.}
\newblock \bibinfo{title}{High-efficiency and broadband on-chip electro-optic frequency comb generators}.
\newblock \emph{\bibinfo{journal}{Nat. Photon.}} \textbf{\bibinfo{volume}{16}}, \bibinfo{pages}{679--685} (\bibinfo{year}{2022}).

\bibitem{wang2024octave}
\bibinfo{author}{Wang, P.-Y.} \emph{et~al.}
\newblock \bibinfo{title}{Octave soliton microcombs in lithium niobate microresonators}.
\newblock \emph{\bibinfo{journal}{Opt. Lett.}} \textbf{\bibinfo{volume}{49}}, \bibinfo{pages}{1729--1732} (\bibinfo{year}{2024}).

\bibitem{song2024octave}
\bibinfo{author}{Song, Y.}, \bibinfo{author}{Hu, Y.}, \bibinfo{author}{Zhu, X.}, \bibinfo{author}{Yang, K.} \& \bibinfo{author}{Lon{\v{c}}ar, M.}
\newblock \bibinfo{title}{Octave-spanning {K}err soliton frequency combs in dispersion-and dissipation-engineered lithium niobate microresonators}.
\newblock \emph{\bibinfo{journal}{Light Sci. Appl.}} \textbf{\bibinfo{volume}{13}}, \bibinfo{pages}{225} (\bibinfo{year}{2024}).

\bibitem{schaufele1966raman}
\bibinfo{author}{Schaufele, R.} \& \bibinfo{author}{Weber, M.}
\newblock \bibinfo{title}{Raman scattering by lithium niobate}.
\newblock \emph{\bibinfo{journal}{Phys. Rev.}} \textbf{\bibinfo{volume}{152}}, \bibinfo{pages}{705} (\bibinfo{year}{1966}).

\bibitem{okawachi2017competition}
\bibinfo{author}{Okawachi, Y.} \emph{et~al.}
\newblock \bibinfo{title}{Competition between {R}aman and {K}err effects in microresonator comb generation}.
\newblock \emph{\bibinfo{journal}{Opt. Lett.}} \textbf{\bibinfo{volume}{42}}, \bibinfo{pages}{2786--2789} (\bibinfo{year}{2017}).

\bibitem{wang2019monolithic}
\bibinfo{author}{Wang, C.} \emph{et~al.}
\newblock \bibinfo{title}{Monolithic lithium niobate photonic circuits for {K}err frequency comb generation and modulation}.
\newblock \emph{\bibinfo{journal}{Nat. commun.}} \textbf{\bibinfo{volume}{10}}, \bibinfo{pages}{978} (\bibinfo{year}{2019}).

\bibitem{yu2020raman}
\bibinfo{author}{Yu, M.} \emph{et~al.}
\newblock \bibinfo{title}{Raman lasing and soliton mode-locking in lithium niobate microresonators}.
\newblock \emph{\bibinfo{journal}{Light Sci. Appl.}} \textbf{\bibinfo{volume}{9}}, \bibinfo{pages}{9} (\bibinfo{year}{2020}).

\bibitem{gong2020photonic}
\bibinfo{author}{Gong, Z.} \emph{et~al.}
\newblock \bibinfo{title}{Photonic dissipation control for $\mathrm{Kerr}$ soliton generation in strongly $\mathrm{Raman}$-active media}.
\newblock \emph{\bibinfo{journal}{Phys. Rev. Lett.}} \textbf{\bibinfo{volume}{125}}, \bibinfo{pages}{183901} (\bibinfo{year}{2020}).

\bibitem{marin2017microresonator}
\bibinfo{author}{Marin-Palomo, P.} \emph{et~al.}
\newblock \bibinfo{title}{Microresonator-based solitons for massively parallel coherent optical communications}.
\newblock \emph{\bibinfo{journal}{Nature}} \textbf{\bibinfo{volume}{546}}, \bibinfo{pages}{274} (\bibinfo{year}{2017}).

\bibitem{suh2016microresonator}
\bibinfo{author}{Suh, M.-G.}, \bibinfo{author}{Yang, Q.-F.}, \bibinfo{author}{Yang, K.~Y.}, \bibinfo{author}{Yi, X.} \& \bibinfo{author}{Vahala, K.~J.}
\newblock \bibinfo{title}{Microresonator soliton dual-comb spectroscopy}.
\newblock \emph{\bibinfo{journal}{Science}} \textbf{\bibinfo{volume}{354}}, \bibinfo{pages}{600--603} (\bibinfo{year}{2016}).

\bibitem{bao2021architecture}
\bibinfo{author}{Bao, C.} \emph{et~al.}
\newblock \bibinfo{title}{Architecture for microcomb-based {GH}z-mid-infrared dual-comb spectroscopy}.
\newblock \emph{\bibinfo{journal}{Nat. Commun.}} \textbf{\bibinfo{volume}{12}}, \bibinfo{pages}{6573} (\bibinfo{year}{2021}).

\bibitem{repelin1999raman}
\bibinfo{author}{Repelin, Y.}, \bibinfo{author}{Husson, E.}, \bibinfo{author}{Bennani, F.} \& \bibinfo{author}{Proust, C.}
\newblock \bibinfo{title}{Raman spectroscopy of lithium niobate and lithium tantalate. {F}orce field calculations}.
\newblock \emph{\bibinfo{journal}{J. Phys. Chem. Solids}} \textbf{\bibinfo{volume}{60}}, \bibinfo{pages}{819--825} (\bibinfo{year}{1999}).

\bibitem{wang2024lithium}
\bibinfo{author}{Wang, C.} \emph{et~al.}
\newblock \bibinfo{title}{Lithium tantalate photonic integrated circuits for volume manufacturing}.
\newblock \emph{\bibinfo{journal}{Nature}} \bibinfo{pages}{1--7} (\bibinfo{year}{2024}).

\bibitem{ji2022compact}
\bibinfo{author}{Ji, X.} \emph{et~al.}
\newblock \bibinfo{title}{Compact, spatial-mode-interaction-free, ultralow-loss, nonlinear photonic integrated circuits}.
\newblock \emph{\bibinfo{journal}{Commun. Phys.}} \textbf{\bibinfo{volume}{5}}, \bibinfo{pages}{84} (\bibinfo{year}{2022}).

\bibitem{he2023high}
\bibinfo{author}{He, Y.} \emph{et~al.}
\newblock \bibinfo{title}{High-speed tunable microwave-rate soliton microcomb}.
\newblock \emph{\bibinfo{journal}{Nat. Commun.}} \textbf{\bibinfo{volume}{14}}, \bibinfo{pages}{3467} (\bibinfo{year}{2023}).

\bibitem{herr2014temporal}
\bibinfo{author}{Herr, T.} \emph{et~al.}
\newblock \bibinfo{title}{Temporal solitons in optical microresonators}.
\newblock \emph{\bibinfo{journal}{Nat. Photon.}} \textbf{\bibinfo{volume}{8}}, \bibinfo{pages}{145--152} (\bibinfo{year}{2014}).

\bibitem{carmon2004dynamical}
\bibinfo{author}{Carmon, T.}, \bibinfo{author}{Yang, L.} \& \bibinfo{author}{Vahala, K.}
\newblock \bibinfo{title}{Dynamical thermal behavior and thermal self-stability of microcavities}.
\newblock \emph{\bibinfo{journal}{Opt. Express}} \textbf{\bibinfo{volume}{12}}, \bibinfo{pages}{4742--4750} (\bibinfo{year}{2004}).

\bibitem{zhang2025fundamental}
\bibinfo{author}{Zhang, J.} \emph{et~al.}
\newblock \bibinfo{title}{Fundamental charge noise in electro-optic photonic integrated circuits}.
\newblock \emph{\bibinfo{journal}{Nat. Phys.}} \bibinfo{pages}{1--8} (\bibinfo{year}{2025}).

\bibitem{yi2015soliton}
\bibinfo{author}{Yi, X.}, \bibinfo{author}{Yang, Q.-F.}, \bibinfo{author}{Yang, K.~Y.}, \bibinfo{author}{Suh, M.-G.} \& \bibinfo{author}{Vahala, K.}
\newblock \bibinfo{title}{Soliton frequency comb at microwave rates in a high-{Q} silica microresonator}.
\newblock \emph{\bibinfo{journal}{Optica}} \textbf{\bibinfo{volume}{2}}, \bibinfo{pages}{1078--1085} (\bibinfo{year}{2015}).

\bibitem{brasch2016photonic}
\bibinfo{author}{Brasch, V.} \emph{et~al.}
\newblock \bibinfo{title}{Photonic chip--based optical frequency comb using soliton {Cherenkov} radiation}.
\newblock \emph{\bibinfo{journal}{Science}} \textbf{\bibinfo{volume}{351}}, \bibinfo{pages}{357--360} (\bibinfo{year}{2016}).

\bibitem{zhou2019soliton}
\bibinfo{author}{Zhou, H.} \emph{et~al.}
\newblock \bibinfo{title}{Soliton bursts and deterministic dissipative {K}err soliton generation in auxiliary-assisted microcavities}.
\newblock \emph{\bibinfo{journal}{Light Sci. Appl.}} \textbf{\bibinfo{volume}{8}}, \bibinfo{pages}{50} (\bibinfo{year}{2019}).

\bibitem{zhang2019sub}
\bibinfo{author}{Zhang, S.} \emph{et~al.}
\newblock \bibinfo{title}{Sub-milliwatt-level microresonator solitons with extended access range using an auxiliary laser}.
\newblock \emph{\bibinfo{journal}{Optica}} \textbf{\bibinfo{volume}{6}}, \bibinfo{pages}{206--212} (\bibinfo{year}{2019}).

\bibitem{lu2019deterministic}
\bibinfo{author}{Lu, Z.} \emph{et~al.}
\newblock \bibinfo{title}{Deterministic generation and switching of dissipative {K}err soliton in a thermally controlled micro-resonator}.
\newblock \emph{\bibinfo{journal}{AIP Adv.}} \textbf{\bibinfo{volume}{9}} (\bibinfo{year}{2019}).

\bibitem{obrzud2017temporal}
\bibinfo{author}{Obrzud, E.}, \bibinfo{author}{Lecomte, S.} \& \bibinfo{author}{Herr, T.}
\newblock \bibinfo{title}{Temporal solitons in microresonators driven by optical pulses}.
\newblock \emph{\bibinfo{journal}{Nat. Photon.}} \textbf{\bibinfo{volume}{11}}, \bibinfo{pages}{600} (\bibinfo{year}{2017}).

\bibitem{li2022efficiency}
\bibinfo{author}{Li, J.} \emph{et~al.}
\newblock \bibinfo{title}{Efficiency of pulse pumped soliton microcombs}.
\newblock \emph{\bibinfo{journal}{Optica}} \textbf{\bibinfo{volume}{9}}, \bibinfo{pages}{231--239} (\bibinfo{year}{2022}).

\bibitem{obrzud2019microphotonic}
\bibinfo{author}{Obrzud, E.} \emph{et~al.}
\newblock \bibinfo{title}{A microphotonic astrocomb}.
\newblock \emph{\bibinfo{journal}{Nat. Photon.}} \textbf{\bibinfo{volume}{13}}, \bibinfo{pages}{31} (\bibinfo{year}{2019}).

\bibitem{anderson2022zero}
\bibinfo{author}{Anderson, M.~H.} \emph{et~al.}
\newblock \bibinfo{title}{Zero dispersion {K}err solitons in optical microresonators}.
\newblock \emph{\bibinfo{journal}{Nat. commun.}} \textbf{\bibinfo{volume}{13}}, \bibinfo{pages}{4764} (\bibinfo{year}{2022}).

\bibitem{xiao2023near}
\bibinfo{author}{Xiao, Z.} \emph{et~al.}
\newblock \bibinfo{title}{Near-zero-dispersion soliton and broadband modulational instability {K}err microcombs in anomalous dispersion}.
\newblock \emph{\bibinfo{journal}{Light Sci. Appl.}} \textbf{\bibinfo{volume}{12}}, \bibinfo{pages}{33} (\bibinfo{year}{2023}).

\bibitem{murata2000optical}
\bibinfo{author}{Murata, H.}, \bibinfo{author}{Morimoto, A.}, \bibinfo{author}{Kobayashi, T.} \& \bibinfo{author}{Yamamoto, S.}
\newblock \bibinfo{title}{Optical pulse generation by electrooptic-modulation method and its application to integrated ultrashort pulse generators}.
\newblock \emph{\bibinfo{journal}{IEEE J. Sel. Top. Quant}} \textbf{\bibinfo{volume}{6}}, \bibinfo{pages}{1325--1331} (\bibinfo{year}{2000}).

\bibitem{zhang2025ultrabroadband}
\bibinfo{author}{Zhang, J.} \emph{et~al.}
\newblock \bibinfo{title}{Ultrabroadband integrated electro-optic frequency comb in lithium tantalate}.
\newblock \emph{\bibinfo{journal}{Nature}} \bibinfo{pages}{1--8} (\bibinfo{year}{2025}).

\bibitem{xue2019super}
\bibinfo{author}{Xue, X.}, \bibinfo{author}{Zheng, X.} \& \bibinfo{author}{Zhou, B.}
\newblock \bibinfo{title}{Super-efficient temporal solitons in mutually coupled optical cavities}.
\newblock \emph{\bibinfo{journal}{Nat. Photon.}} \textbf{\bibinfo{volume}{13}}, \bibinfo{pages}{616--622} (\bibinfo{year}{2019}).

\bibitem{helgason2023surpassing}
\bibinfo{author}{Helgason, {\'O}.~B.} \emph{et~al.}
\newblock \bibinfo{title}{Surpassing the nonlinear conversion efficiency of soliton microcombs}.
\newblock \emph{\bibinfo{journal}{Nat. Photon.}} \textbf{\bibinfo{volume}{17}}, \bibinfo{pages}{992--999} (\bibinfo{year}{2023}).

\bibitem{yang2024efficient}
\bibinfo{author}{Yang, Q.-F.}, \bibinfo{author}{Hu, Y.}, \bibinfo{author}{Torres-Company, V.} \& \bibinfo{author}{Vahala, K.}
\newblock \bibinfo{title}{Efficient microresonator frequency combs}.
\newblock \emph{\bibinfo{journal}{eLight}} \textbf{\bibinfo{volume}{4}}, \bibinfo{pages}{18} (\bibinfo{year}{2024}).

\bibitem{yu2022integrated}
\bibinfo{author}{Yu, M.} \emph{et~al.}
\newblock \bibinfo{title}{Integrated femtosecond pulse generator on thin-film lithium niobate}.
\newblock \emph{\bibinfo{journal}{Nature}} \textbf{\bibinfo{volume}{612}}, \bibinfo{pages}{252--258} (\bibinfo{year}{2022}).

\bibitem{gao2023compact}
\bibinfo{author}{Gao, Y.} \emph{et~al.}
\newblock \bibinfo{title}{Compact lithium niobate microring resonators in the ultrahigh {Q}/{V} regime}.
\newblock \emph{\bibinfo{journal}{Opt. Lett.}} \textbf{\bibinfo{volume}{48}}, \bibinfo{pages}{3949--3952} (\bibinfo{year}{2023}).

\bibitem{xue2015mode}
\bibinfo{author}{Xue, X.} \emph{et~al.}
\newblock \bibinfo{title}{Mode-locked dark pulse {K}err combs in normal-dispersion microresonators}.
\newblock \emph{\bibinfo{journal}{Nat. Photon.}} \textbf{\bibinfo{volume}{9}}, \bibinfo{pages}{594--600} (\bibinfo{year}{2015}).

\bibitem{xue2023dispersion}
\bibinfo{author}{Xue, X.} \emph{et~al.}
\newblock \bibinfo{title}{Dispersion-less {K}err solitons in spectrally confined optical cavities}.
\newblock \emph{\bibinfo{journal}{Light Sci. Appl.}} \textbf{\bibinfo{volume}{12}}, \bibinfo{pages}{19} (\bibinfo{year}{2023}).

\bibitem{wu2024visible}
\bibinfo{author}{Wu, T.-H.} \emph{et~al.}
\newblock \bibinfo{title}{Visible-to-ultraviolet frequency comb generation in lithium niobate nanophotonic waveguides}.
\newblock \emph{\bibinfo{journal}{Nat. Photon.}} \textbf{\bibinfo{volume}{18}}, \bibinfo{pages}{218--223} (\bibinfo{year}{2024}).

\bibitem{lu2024emerging}
\bibinfo{author}{Lu, X.} \emph{et~al.}
\newblock \bibinfo{title}{Emerging integrated laser technologies in the visible and short near-infrared regimes}.
\newblock \emph{\bibinfo{journal}{Nat. Photon.}} \textbf{\bibinfo{volume}{18}}, \bibinfo{pages}{1010--1023} (\bibinfo{year}{2024}).

\bibitem{isichenko2023photonic}
\bibinfo{author}{Isichenko, A.} \emph{et~al.}
\newblock \bibinfo{title}{Photonic integrated beam delivery for a rubidium 3{D} magneto-optical trap}.
\newblock \emph{\bibinfo{journal}{Nat. Commun.}} \textbf{\bibinfo{volume}{14}}, \bibinfo{pages}{3080} (\bibinfo{year}{2023}).

\bibitem{Hummon:18}
\bibinfo{author}{Hummon, M.~T.} \emph{et~al.}
\newblock \bibinfo{title}{Photonic chip for laser stabilization to an atomic vapor with $10^{-11}$ instability}.
\newblock \emph{\bibinfo{journal}{Optica}} \textbf{\bibinfo{volume}{5}}, \bibinfo{pages}{443--449} (\bibinfo{year}{2018}).

\bibitem{stern2013nanoscale}
\bibinfo{author}{Stern, L.}, \bibinfo{author}{Desiatov, B.}, \bibinfo{author}{Goykhman, I.} \& \bibinfo{author}{Levy, U.}
\newblock \bibinfo{title}{Nanoscale light--matter interactions in atomic cladding waveguides}.
\newblock \emph{\bibinfo{journal}{Nat. Commun.}} \textbf{\bibinfo{volume}{4}}, \bibinfo{pages}{1548} (\bibinfo{year}{2013}).

\end{thebibliography}

\medskip
\noindent\textbf{Declarations}

\medskip
\noindent\textbf{Ethics approval and consent to participate}

\begin{footnotesize}
\noindent Not applicable.
\end{footnotesize}

\medskip
\noindent\textbf{Consent for publication}

\begin{footnotesize}
\noindent Not applicable.
\end{footnotesize}

\medskip
\noindent\textbf{Data availability}

\begin{footnotesize}
\noindent The data that support the plots within this paper and other findings of this study are available from the corresponding author upon reasonable request.
\end{footnotesize}

\medskip

\noindent\textbf{Code availability}

\begin{footnotesize}
\noindent The codes that support the findings of this study are available from the corresponding authors upon reasonable request. 
\end{footnotesize}

\medskip

\noindent\textbf{Acknowledgments}

\begin{footnotesize}
\noindent 
The authors thank Yaowen Hu for helpful discussions.
\end{footnotesize}

\medskip
\noindent\textbf{Funding}

\begin{footnotesize}
\noindent
This work was supported by Beijing Natural Science Foundation (Z210004), National Natural Science Foundation of China (92150108), the high-performance computing Platform of Peking University, and the Advanced Photonic Integrated Center (APIC) of State Key Laboratory of Advanced Optical Communication System and Networks.
\end{footnotesize}

\medskip

\noindent\textbf{Author contributions} 

\begin{footnotesize}
\noindent 
The experiment was conceived by B.N., X.L., C.Y., and Q.-F.Y. Measurements were performed by B.N., X.L., K.Z., and Z.W. The data was analyzed by B.N. The Raman spectroscopy was performed by G.Z., B.N., X.L., and C.Y. The device was designed by X.L. and fabricated by X.L, C.Y, and R.M. All authors contributed to writing the manuscript.
\end{footnotesize}

\vspace{6pt}
\noindent\textbf{Competing interests}

\begin{footnotesize}
\noindent The authors declare no competing interests.
\end{footnotesize}

\medskip

\noindent\textbf{Additional information}

\begin{footnotesize}
\noindent Supplementary information is available for this paper.

\medskip

\noindent {\bf Correspondence and requests for materials} should be addressed to Q.-F.Y.

\newpage
\begin{figure*}[h!]
\setcounter{figure}{0}
\centering
    \includegraphics[scale = 1.0]{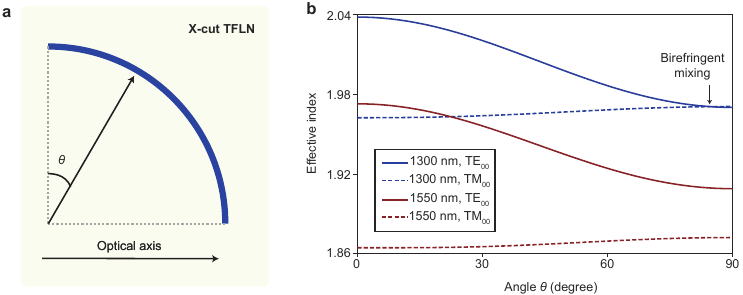}
    \renewcommand{\figurename}
    {\textbf{Extended Data Fig.}}
    \caption{ { {\bf Birefringence induced mode mixing.}}
    \textbf{a,} Illustration of the curve angle $\theta$, defined as the angle between the optical axis and the direction of light propagation.
    \textbf{b,} Numerical simulation of the effective indices for the TE$_{00}$ and TM$_{00}$ fundamental modes as a function of curve angle $\theta$. Birefringent mixing between TE$_{00}$ and TM$_{00}$ modes is observed at shorter wavelengths (1300 nm) but is absent at longer wavelengths (1550 nm).  }
    \label{fig:ExtFig1}
\end{figure*}

\begin{figure*}[h!]
\setcounter{figure}{1}
\centering
    \includegraphics[scale = 1.0]{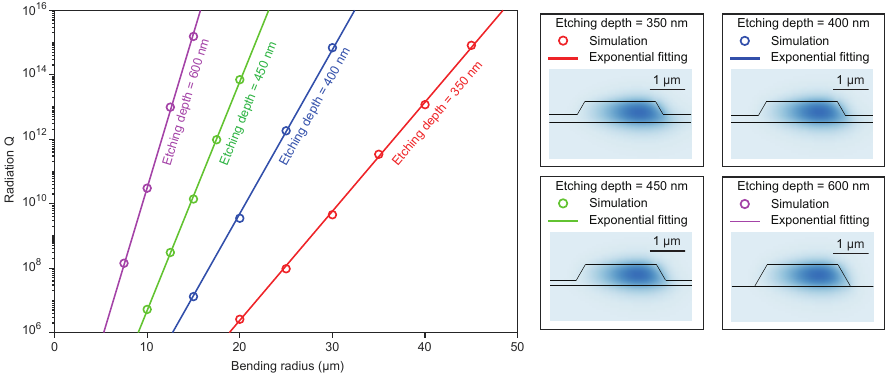}
    \renewcommand{\figurename}
    {\textbf{Extended Data Fig.}}
    \caption{{ {\bf Simulation of bending induced radiation loss.} }
    {The radiation Q factor of the TE fundamental mode is plotted as a function of bending radius for waveguides with different etching depths. The thickness of TFLN is 600 nm, and the waveguide width is 2 \textmu m. The radiation Q factor is significantly enhanced with increasing etching depth.}}
    \label{fig:ExtFig2}
\end{figure*}

\end{footnotesize}

\smallskip

\end{document}